\documentclass[3p]{elsarticle}

\usepackage{hyperref}
\usepackage{siunitx}
\usepackage{booktabs}
\usepackage{amsmath}
\usepackage{amssymb}

\journal{Journal of Quantitative Spectroscopy and Radiative Transfer}

\def\be{\begin{equation}}
\def\ee{\end{equation}}
\def\bes{\begin{equation*}}
\def\ees{\end{equation*}}

\begin{document}

\begin{frontmatter}

\title{Computational assessment of an effective-sphere model for characterizing colloidal fractal aggregates with holographic microscopy}

\author[1]{Jerome Fung\corref{cor1}}
\ead{jfung@ithaca.edu}
\address[1]{Department of Physics \& Astronomy, Ithaca College, 953 Danby Road, Ithaca, NY 14850}

\author[2]{Samantha Hoang}
\ead{shoang@wellesley.edu}
\address[2]{Department of Physics, Wellesley College, 106 Central Street, Wellesley, MA 02481}

\cortext[cor1]{Corresponding author}

\begin{abstract}
We perform simulations to evaluate a recent experimental technique for using in-line holographic microscopy and an effective-sphere
model to measure the population-averaged fractal dimension $D_f$ of an ensemble of colloidal fractal aggregates.
In this technique, models based on Lorenz-Mie scattering by a uniform sphere are fit to digital holograms of a population of fractal aggregates to determine the effective refractive indices $n_{eff}$ and effective radii $a_{eff}$ of the aggregates.
A scaling relationship between $n_{eff}$ and $a_{eff}$ based on the Maxwell Garnett effective-medium theory then determines $D_f$.
Here we use a multisphere superposition code to calculate the exact holograms produced by aggregates with tunable fractal dimensions $D_f$.
We show that $n_{eff}$ and $a_{eff}$ become less sensitive to the aggregate orientation as $D_f$ increases.
We also show that the Maxwell Garnett scaling relationship 
correctly determines $D_f$ to within 10.5\% when multiple scattering is negligible and the population-averaged coefficient of determination $\langle R^2\rangle_p > 0.6$, indicating that the holograms are well-described by the effective-sphere model. 
\end{abstract}


\end{frontmatter}


\section{Introduction}
Fractal-like aggregates of small particles commonly arise from both natural and artificial processes.
Examples include soot \cite{berry_optics_1986, bond_light_2006}, protein aggregates formed by heating \cite{feder_scaling_1984, hagiwara_fractal_1996, wu_stability_2012}, and aggregates of polystyrene spheres formed by salting out a suspension  \cite{pusey_measurement_1987, zhou_light-scattering_1991}. 
Experimental studies on fractal aggregates have frequently focused on determining their fractal dimension $D_f$, which characterizes their structure and strongly affects how they interact with light \cite{sorensen_light_2001}.
Direct imaging of individual aggregates using scanning or transmission electron microscopy \cite{chakrabarty_low_2009} can allow detailed examination of their structure and the measurement of $D_f$.
However, in applications such as the modeling of radiative transfer by an ensemble of particles, the fractal dimension of a \emph{population} of aggregates, rather than the structure of a single aggregate, is the quantity of interest.
Determining population-averaged fractal dimensions requires the time-consuming imaging of many aggregates; moreover, electron microscopy cannot be performed \emph{in situ}.
Macroscopic light scattering can determine fractal dimensions that are averaged over many aggregates. 
However, bulk light scattering techniques can be affected by polydispersity, and interpreting the results may require strong assumptions to be made about the scatterers \cite{sorensen_light_2001}.

Recently, Wang and co-workers pioneered the use of in-line holographic microscopy for characterizing non-absorbing fractal aggregates \cite{wang_holographic_2016}. 
In holographic microscopy (also known as digital holographic microscopy or holographic video microscopy), light scattered from an aggregate of interest interferes with unscattered incident light to form an interference pattern, or hologram. 
Rather than using reconstruction techniques to produce an image of the aggregate \cite{holler_two-dimensional_2018}, 
Wang \emph{et al.} analyzed holograms of fractal aggregates by fitting effective-sphere models based on Lorenz-Mie scattering
by a uniform sphere. 
From their fits, they determined each aggregate's effective radius $a_{eff}$ and effective refractive index $n_{eff}$.
Using the Maxwell Garnett effective-medium theory, Wang \emph{et al.} showed that the relationship between $a_{eff}$ and $n_{eff}$ for aggregates of different sizes within a population depends on $D_f$. 
Wang \emph{et al.} applied this approach to colloidal aggregates formed \emph{in situ} from polystyrene spheres, bovine pancreas insulin, and bovine serum albumin as they flowed through a microfluidic device.

While the $D_f$ values Wang \emph{et al.} determined for their aggregates were consistent with their expectations, questions remain about the limits of validity of their approach.
In this work, we computationally test their approach of using holographic microscopy and an effective-sphere model to characterize the fractal dimension $D_f$ of a population of aggregates.
We generate aggregates with a known, tunable $D_f$ that are composed of monodisperse, non-absorbing spheres.
We use an exact multisphere superposition code to calculate the holograms that those aggregates would propose.
We then fit effective-sphere models to the exact holograms, determine $n_{eff}$ and $a_{eff}$ for each aggregate, and infer $D_f$ from the distribution of $n_{eff}$ and $a_{eff}$ for a population of aggregates.
We compare the values of $D_f$ inferred from holographic microscopy to the input $D_f$ values for aggregates composed of primary spheres of different sizes.
Our results suggest criteria that may help experimenters determine whether $D_f$ values obtained using the effective-sphere approach of Wang \emph{et al.} are reliable.

\section{Theory}
\subsection{The fractal dimension of a fractal aggregate}
The geometry of a fractal-like aggregate is characterized by two parameters: the fractal dimension $D_f$ and the fractal prefactor $k_f$. 
An aggregate composed of $N$ spheres each of radius $a_0$ (known as \emph{primary spheres}) satisfies the geometric relationship
\be \label{eq:df_defn}
N = k_f\left(\frac{R_g}{a_0} \right)^{D_f}.
\ee
Here, $R_g$ is the radius of gyration of the aggregate. In general, as $D_f$ increases, aggregates become increasingly close-packed.
A straight linear chain of particles has $D_f=1$, while a three-dimensional crystalline array of particles has $D_f=3$.
We focus here on the characterization of $D_f$, as do Wang \emph{et al.} \cite{wang_holographic_2016}.

\subsection{Holographic microscopy}
\begin{figure}[]
\centering
\includegraphics{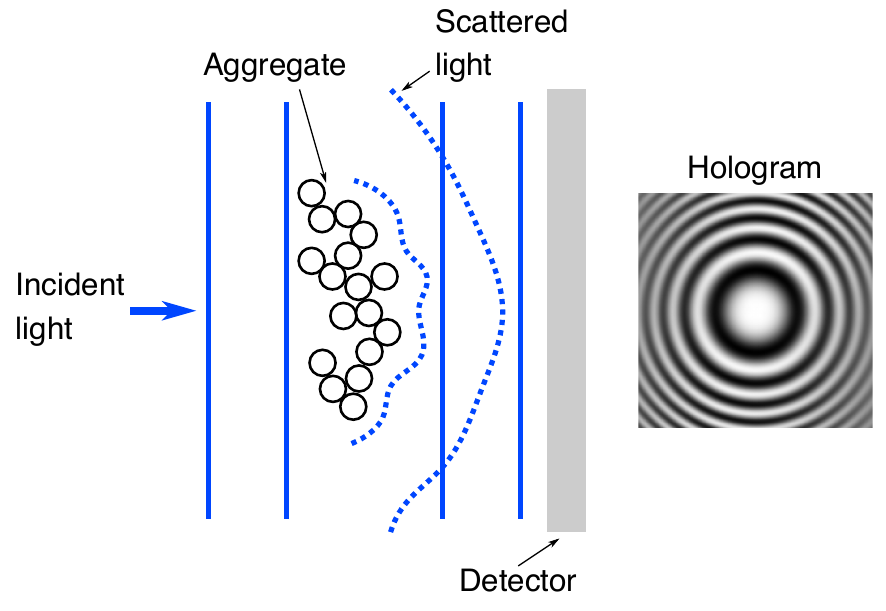}
\caption{\label{fig:schematic} Schematic illustration of in-line holographic microscopy. The hologram recorded on a detector is an interference pattern formed by an incident plane wave and the wave scattered by an object of interest.}
\end{figure}

In in-line holographic microscopy, 
the recorded hologram is an interference pattern formed by the scattered wave from an object of interest and the unscattered portion of the incident wave (Fig.~\ref{fig:schematic}).
Following Lee and co-workers, we model the formation of a normalized digital hologram using \cite{lee_characterizing_2007}
\be \label{eq:dhm}
I_{hol} = 1 + 2 \alpha \operatorname{Re} \left\{ \vec{E}_{s} \cdot \hat{\mathbf{\epsilon}}   \right\} + \alpha^2 | \vec{E}_{s} |^2.
\ee
Here $I_{hol}$ is the measured hologram intensity at a location on the detector, $\vec{E}_{s}$ is the complex scattered electric field, $\hat{\mathbf{\epsilon}}$ is a unit vector in the direction of the polarization of the incident field, $\alpha$ is a scaling parameter that depends on details of the optical system, and $\operatorname{Re}$ denotes the real part.
The first term on the right side of Eq.~(\ref{eq:dhm}) is proportional to the intensity of the incident field prior to normalization, the second term encodes the amplitude and phase of the scattered wave, and the third term is proportional to the intensity of the scattered field. 
A hologram thus contains information about the position of the scatterers (via the variation of the phase of $\vec{E}_{s}$ on the detector) and about the optical properties of the scatterers (via the angular dependence of $\vec{E}_s$).
Once $\vec{E}_s$ can be calculated for an object, Eq.~(\ref{eq:dhm}) can be used to compute the hologram formed by that object.

For dilute samples in which the scatterers are comparable in size to the wavelength of light and are well-separated, information about the scatterers can be retrieved from experimentally recorded holograms by fitting a computational model based on Eq.~(\ref{eq:dhm}). 
Both the three-dimensional positions of the scatterers as well as their optical properties (such as their size and refractive index) can be determined.
This has been demonstrated for isolated spheres \cite{lee_characterizing_2007, shpaisman_holographic_2012}, clusters of spheres \cite{fung_imaging_2012}, and cylinders \cite{wang_using_2014}. 
Here we only consider the optical properties of the scatterers.

\subsection{Modeling fractal aggregates using effective-medium theory}
Wang \emph{et al.}~use the Maxwell Garnett effective-medium theory \cite{garnett_j._c._maxwell_xii._1904, markel_introduction_2016} to predict $n_{eff}$ and $a_{eff}$ for fractal aggregates modeled as uniform spheres \cite{wang_holographic_2016}. 
They assume that an aggregate with radius of gyration $R_g$ can be modeled as a uniform sphere of radius $R_g$. 
While a sphere of radius $R_g$ does not necessarily enclose the aggregate, particularly for low $D_f$ aggregates,
they define a volume fraction $\phi$ equal to the ratio between the volume of a sphere of radius $R_g$ and the total volume of the $N$ primary particles of radius $a_0$ in the aggregate:
\be 
\phi \equiv \frac{N\cdot \frac{4}{3}\pi a_0^3}{\frac{4}{3}\pi R_g^3}.
\ee
Using Eq.~(\ref{eq:df_defn}) to eliminate $N$ and assuming that $R_g=a_{eff}$ leads to
\be \label{eq:vol_frac}
\phi = k_f \left( \frac{a_{eff}}{a_0}  \right)^{D_f-3}.
\ee
The Maxwell Garnett effective-medium theory relates the Lorentz-Lorenz factors $L(n)$ of the refractive indices $n_{eff}$ of the effective spheres and $n_0$ of the primary particles:
\be \label{eq:MG}
L(n_{eff}) = \phi L(n_0)
\ee
where
\be \label{eq:lorentz_lorenz}
L(n) = \frac{n^2-n_{med}^2}{n^2+2n_{med}^2}.
\ee
Here $n_{med}$ is the refractive index of the surrounding medium.
Combining Eqs.~(\ref{eq:vol_frac}) and (\ref{eq:MG}) leads to
\be \label{eq:main_scaling}
\ln\left(  \frac{L(n_{eff})}{L(n_0)}  \right) = (D_f-3) \ln\left( \frac{a_{eff}}{a_0}  \right) + \ln k_f.
\ee
This is the scaling relationship between $n_{eff}$ and $a_{eff}$ from which Wang \emph{et al.} determine $D_f$.

\section{Simulation methods}
\subsection{Generation of fractal aggregates}

\begin{table}[]
\centering
\begin{tabular}{l l }\toprule
Fractal dimension $D_f$ & Fractal prefactor $k_f$ \\ \midrule
1.1 & 1.6 \\
1.3 & 1.6 \\
1.5 & 1.4 \\
1.7 & 1.4 \\
1.9 & 1.2 \\
2.1 & 1.2 \\
2.3 & 1.0 \\
2.5 & 1.0 \\ \bottomrule
\end{tabular}
\caption{\label{tab:geometries} Fractal dimensions and prefactors for aggregates considered in this work.}
\end{table}

In this work, we consider fractal aggregates with the $D_f$ and $k_f$ values shown in Table \ref{tab:geometries}.
We generate aggregates using the algorithms of Filippov \emph{et al.} \cite{filippov_fractal-like_2000} as implemented in the software package FLAGE \cite{skorupski_fast_2014}.
There are two main approaches for generating fractal aggregate geometries in which both the number of particles $N$ and the fractal dimension $D_f$ are tunable.
One approach, known as particle-cluster aggregation or the sequential algorithm, involves adding particles one at a time to a growing aggregate such that Eq.~(\ref{eq:df_defn}) is always satisfied \cite{mackowski_electrostatics_1995}. 
In the other approach, cluster-cluster aggregation, larger aggregates are created by repeatedly merging smaller sub-clusters in a manner that satisfies Eq.~(\ref{eq:df_defn}) \cite{thouy_cluster-cluster_1994}.
While either approach generates aggregates whose radius of gyration is well-described by Eq.~(\ref{eq:df_defn}), the cluster-cluster approach tends to produce clusters whose pair correlation function scales as expected \cite{filippov_fractal-like_2000}.
We therefore use a cluster-cluster aggregation algorithm to generate all aggregates except for those with $D_f=1.1$, for which we use a particle-cluster algorithm. 
Aggregates with $D_f=1.1$ are so nearly linear that very few cluster-cluster mergers are possible, which causes the cluster-cluster algorithm to seldom converge.

\subsection{Calculation of exact holograms}

Throughout our work, except in Sec.~\ref{subsec:high_df}, we assume that the primary particles are monodisperse, non-absorbing spheres whose refractive index is $n_0 = 1.59$, and we assume that the spheres are surrounded by a medium with refractive index $n_{med} = 1.33$. 
These values correspond to polystyrene spheres in water, one of the experimental systems studied by Wang \emph{et al.}
We consider three different values for the radius $a_0$ of the primary particles: \SI{5}{\nano\meter}, \SI{20}{\nano\meter}, and \SI{80}{\nano\meter}.
We also assume that the aggregates are illuminated by incident light with a vacuum wavelength of $\lambda_0=\SI{447}{\nano\meter}$ that is linearly polarized in the horizontal direction.
We assume that our detector is an array of $180\times180$ pixels with a pixel size of \SI{0.1}{\micro\meter}. 
We place all aggregates so that their centers of mass are located \SI{30}{\micro\meter} from the detector along the optical axis and are centered on the detector. 
While a microscope objective lens usually images holograms onto a camera (typically with a pixel size of $\sim\SI{10}{\micro\meter}$) in experiments, we do not explicitly model the lens except for its effect on the pixel size.
To model the aggregates being randomly tumbled as in the experiments of Wang \emph{et al.}, we randomly rotate aggregates about their centers of mass using an algorithm due to Brannon \cite{brannon_rotation_2018}. 
Two unit vectors uniformly distributed on the unit sphere are generated and then orthogonalized using the Graham-Schmidt procedure. The cross product of the resulting vectors then yields a third orthonormal vector to define a rotation matrix.

In order to calculate the hologram formed by a given aggregate using Eq.~(\ref{eq:dhm}), it is necessary to calculate the electric field scattered by the aggregate.
We use the multisphere superposition code SCSMFO1B by Mackowski and Mishchenko \cite{mackowski_calculation_1996}.
This code calculates the exact scattered field (up to series truncation) from a cluster of arbitrary, non-overlapping spheres as a series expansion in
vector spherical harmonics. 
The SCSMFO1B code is wrapped into the open-source package HoloPy (version 3.2.1), which allows for convenient generation and analysis of holograms \cite{barkley_holographic_2018}.

\subsection{Fitting efffective-sphere models}

We also use HoloPy to fit effective-sphere models to exact holograms. We use the Levenberg-Marquardt solver \texttt{nmpfit} (as wrapped in HoloPy) to
fit a model based on the Lorenz-Mie scattering solution \cite{lee_characterizing_2007}. 
In fitting the model, we vary the particle refractive index $n_{eff}$, the particle radius $a_{eff}$, 
the $x$, $y$, and $z$ coordinates of the sphere center, and the scaling parameter $\alpha$ that regulates the amplitude of the scattered electric field.
Nonlinear least squares algorithms (like the Levenberg-Marquardt algorithm) require initial guesses for the parameter values. 
We provide initial guesses for $n_{eff}$ and $a_{eff}$ using the values that are predicted by Eq.~(\ref{eq:main_scaling}). 
If necessary, the need to provide initial guesses could be avoided by using Bayesian inference or machine learning techniques \cite{dimiduk_bayesian_2016, yevick_machine-learning_2014}.
We reject any fits where the minimizer does not return parameter values that differ from the initial guess values, which would indicate that the minimizer did not converge.

We assess the goodness of fit of the effective-sphere models to the exact holograms using the $R^2$ coefficient of determination, a dimensionless parameter given by \cite{fung_imaging_2012}
\be \label{eq:rsquared}
R^2 = 1 - \frac{\sum(I_{holo} - I_{model})^2}{\sum (I_{holo} - \bar{I}_{holo})^2}.
\ee
Here, $I_{holo}$ denotes the pixel values of an exact hologram, $I_{model}$ denotes the pixel values of the best-fit model, $\bar{I}_{holo}$ denotes the mean value of the exact hologram pixels, and the sums are taken over all hologram pixels. 
When the model matches the exact hologram, $R^2=1$.

\section{Results and Discussion}

\subsection{Exact hologram calculations and effective-sphere model fits}
\begin{figure}[]
\includegraphics[width=\textwidth]{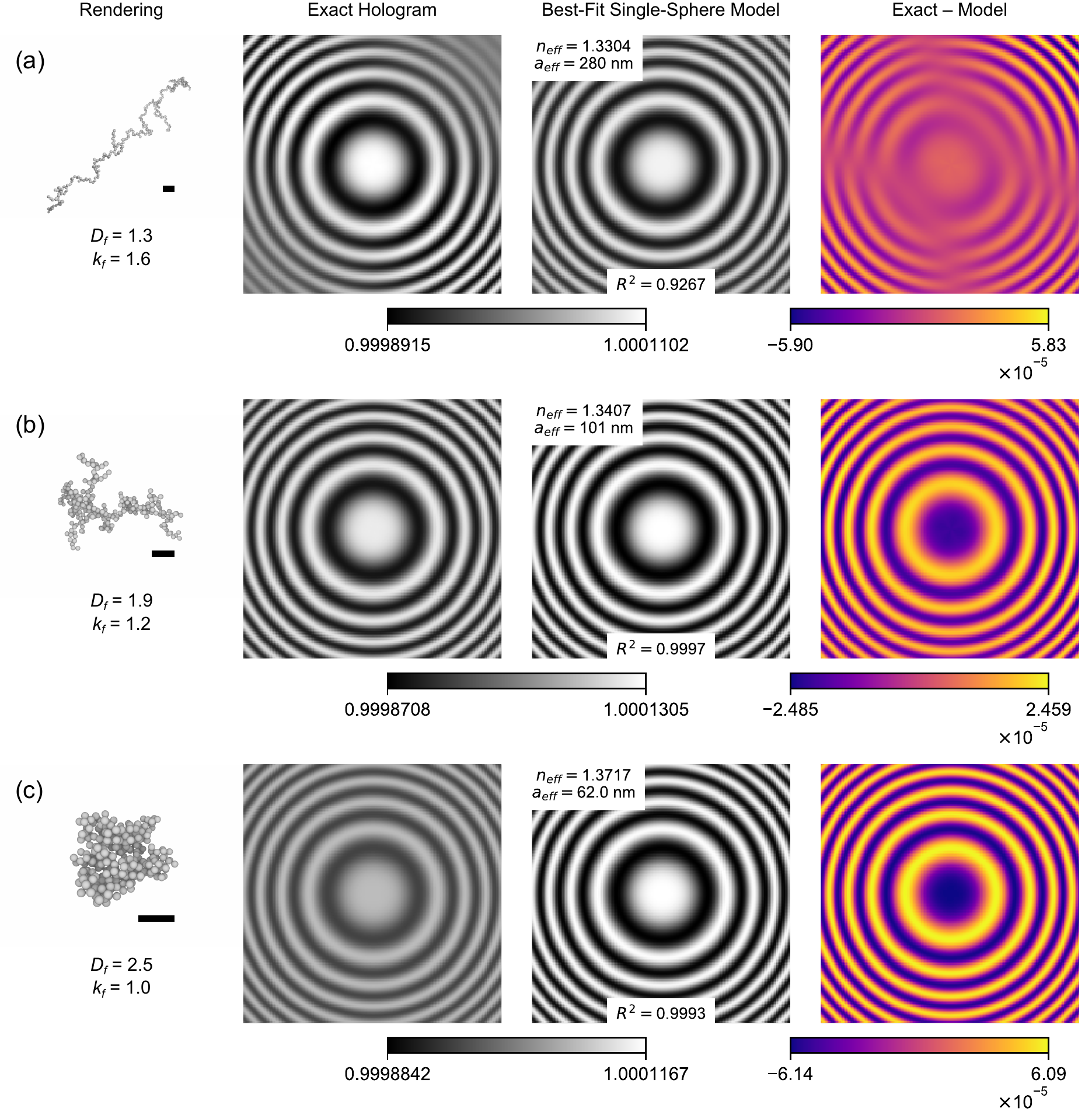}
\caption{\label{fig:sample_holograms} Holograms produced by aggregates composed of $N=300$ primary spheres with different fractal dimensions. 
(a) $D_f = 1.3$, $k_f=1.6$. (b) $D_f=1.9$, $k_f=1.2$. (c) $D_f=2.5$, $k_f=1.0$.
A rendering of the aggregate (in which the incident light propagates out of the page), the exact hologram, the best-fit effective-sphere hologram, 
and the difference between the exact and best-fit holograms are shown for each case. 
For each fit, the effective refractive index $n_{eff}$, the effective radius $a_{eff}$, and the $R^2$ coefficient of determination are shown.
The primary spheres are 5-\si{\nano\meter}-radius polystyrene spheres in water. 
Scale bars: 50 nm.
}
\end{figure}

Figure \ref{fig:sample_holograms} shows renderings, exact holograms, best-fit effective-sphere models, and residuals for three aggregates with different fractal dimensions each consisting of $N=300$ primary spheres.
The renderings show how the aggregates are nearly linear when $D_f=1.3$ but are much more tightly packed when $D_f = 2.5$.
The holograms of these three aggregates are well-described by the effective-sphere model, as indicated by the values of $R^2 > 0.9$. 
The residuals show subtler differences between the exact and effective-sphere holograms. 
In particular, the elongated nature of the aggregate with $D_f=1.3$ in Figure \ref{fig:sample_holograms}(a) leads to the diagonal bands in the residuals.

\subsection{Orientation dependence of $n_{eff}$ and $a_{eff}$}
In the experiments of Wang \textit{et al.}, the aggregates tumble as they flow through a microfluidic channel \cite{wang_holographic_2016}. 
Each aggregate is typically imaged in several orientations, and the resulting values of $n_{eff}$ and $a_{eff}$ are averaged together. 
It is therefore worthwhile to assess how sensitive $n_{eff}$ and $a_{eff}$ are to the aggregate orientation.

\begin{figure}[]
\includegraphics[width=\textwidth]{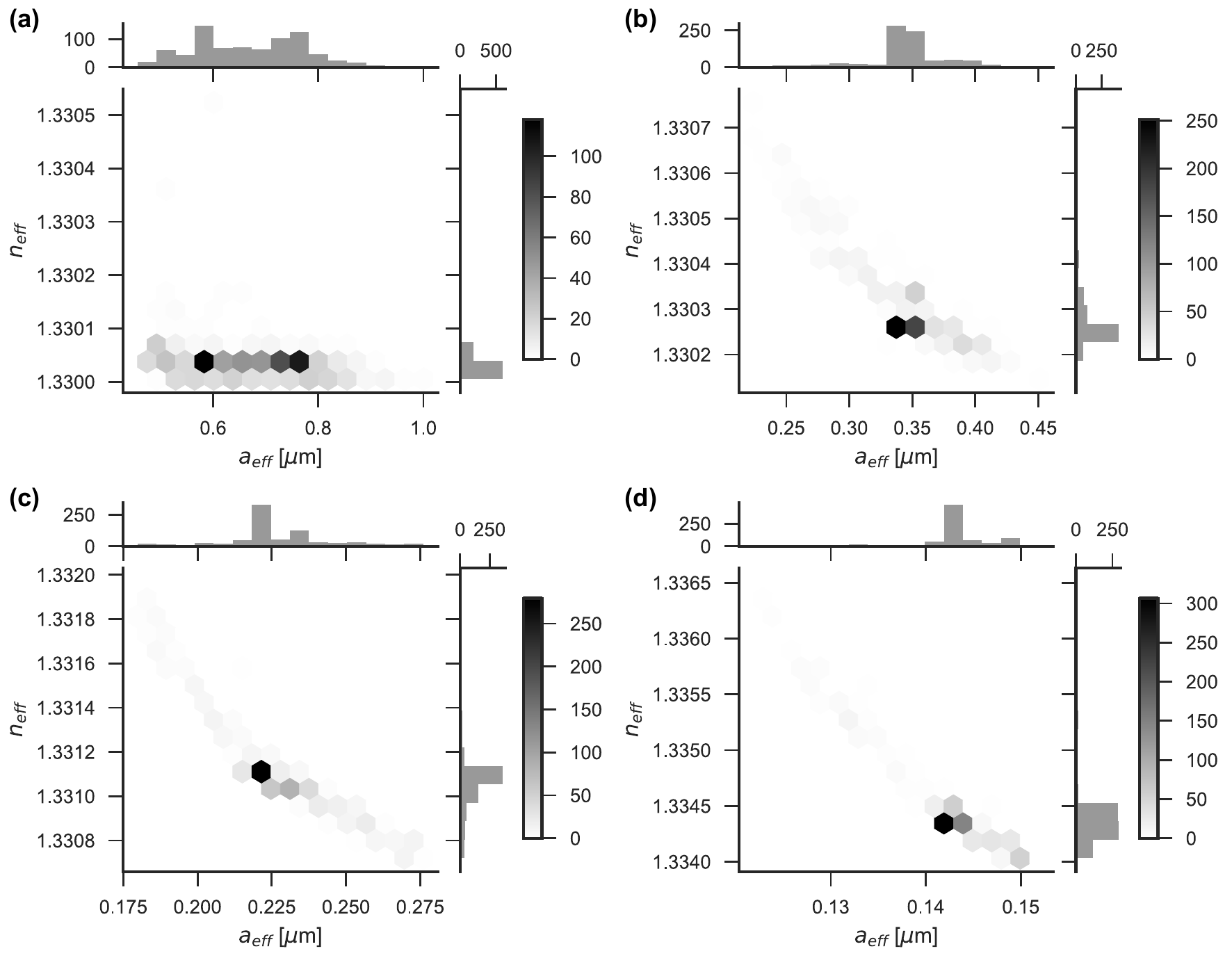}
\caption{\label{fig:rot_dists_1}
Joint and marginal histograms of effective-sphere refractive indices $n_{eff}$ and radii $a_{eff}$ for aggregates of 5-nm-radius primary spheres with (a) $D_f = 1.1$, (b) $D_f = 1.3$, (c) $D_f=1.5$, and (d) $D_f=1.7$. Each plot shows results for 800 aggregate orientations.
}
\end{figure}

\begin{figure}[]
\includegraphics[width=\textwidth]{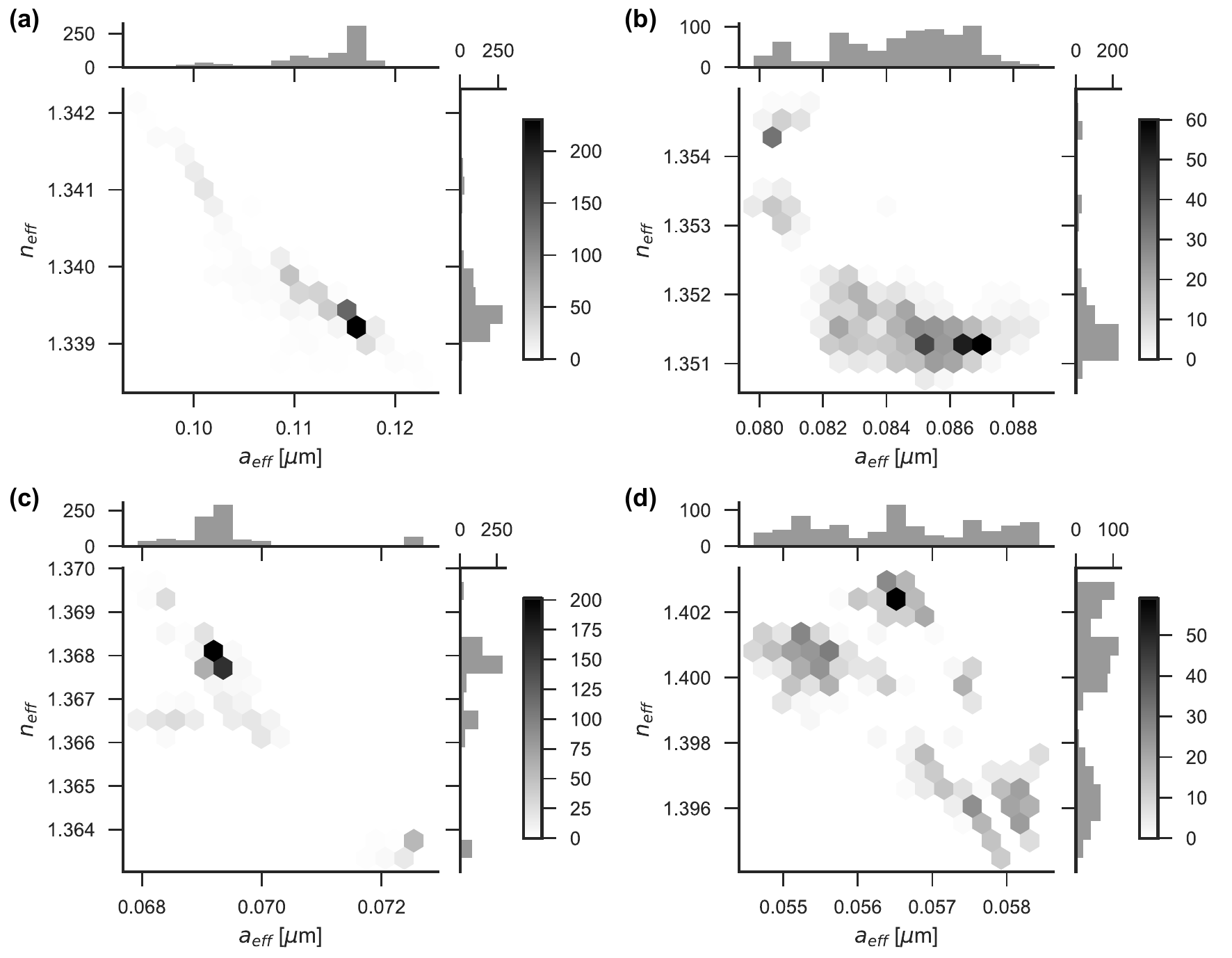}
\caption{\label{fig:rot_dists_2}
Joint and marginal histograms of effective-sphere refractive indices $n_{eff}$ and radii $a_{eff}$ for aggregates of 5-nm-radius primary spheres with (a) $D_f = 1.9$, (b) $D_f = 2.1$, (c) $D_f=2.3$, and (d) $D_f=2.5$. Each plot shows results for 800 aggregate orientations.
}
\end{figure}

To study this question, we generate one aggregate with $N=300$ primary spheres of radius $a_0 = \SI{5}{\nano\meter}$ for each of the fractal dimensions in Table \ref{tab:geometries}. 
We calculate exact holograms for 800 randomly-chosen orientations for each aggregate, determine $n_{eff}$ and $a_{eff}$ from each hologram, and compile the resulting distributions of $n_{eff}$ and $a_{eff}$ in Figures \ref{fig:rot_dists_1} and \ref{fig:rot_dists_2}.

\begin{figure}[]
\centering
\includegraphics[]{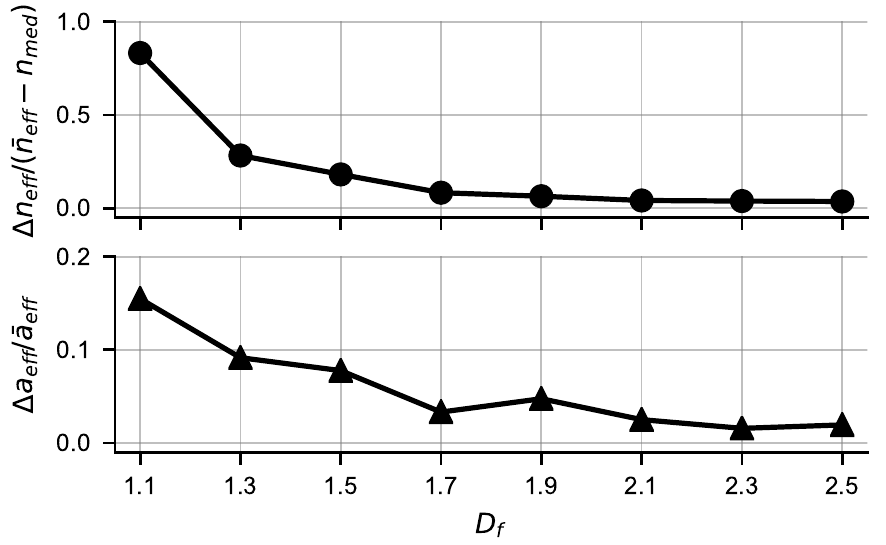}
\caption{ \label{fig:rot_summary}
Normalized sample standard deviations from marginal distributions of $n_{eff}$ and $a_{eff}$ values for aggregates with different fractal dimensions $D_f$.
}
\end{figure}

We make two principal observations about Figures \ref{fig:rot_dists_1} and \ref{fig:rot_dists_2}.
First, $n_{eff}$ and $a_{eff}$ are anti-correlated. This is consistent with experimental observations for single spheres  \cite{lee_characterizing_2007}. 
This anti-correlation results from the product of  the sphere radius and the sphere refractive index frequently appearing in the Lorenz-Mie solution.
Second, as $D_f$ increases, the distributions of $n_{eff}$ and $a_{eff}$ generally become narrower. 
While the marginal distributions are not always monomodal (such as the $a_{eff}$ distribution for $D_f=2.3$ in Fig.~\ref{fig:rot_dists_2}(b)), it is still possible to calculate normalized sample standard deviations (Fig.~\ref{fig:rot_summary}). 
For $n_{eff}$, we normalize the  sample standard deviation $\Delta n_{eff}$ by the difference between the sample mean $\bar{n}_{eff}$ and the medium index $n_{med}$.
For $a_{eff}$, we normalize the sample standard deviation $\Delta a_{eff}$ by the sample mean $\bar{a}_{eff}$.
The normalized sample standard deviations decrease as $D_f$ increases because the morphology of the aggregates becomes increasingly compact.
For elongated low-$D_f$ aggregates, the size of the geometric projection of the aggregate onto the hologram plane (which is correlated with $a_{eff}$) strongly depends on the orientation of the aggregate relative to the optical axis.
For more compact high-$D_f$ aggregates, the projections are approximately the same no matter how the aggregates are oriented relative to the optical axis. 
Consequently, $a_{eff}$ and $n_{eff}$ are much less sensitive to the aggregate orientation.

\subsection{Determination of $D_f$ using holographic microscopy}
We next consider whether the values of $D_f$ determined using holographic microscopy and the scaling relation between $n_{eff}$ and $a_{eff}$ [Eq.~(\ref{eq:main_scaling})] are consistent with the known input values.
For each of the fractal dimensions in Table \ref{tab:geometries}, we calculate exact holograms for 225 aggregates consisting of between $N=60$ and $N=600$ primary particles.
We calculate holograms for five random orientations of each aggregate.
We then determine $n_{eff}$ and $a_{eff}$ for each hologram by fitting an effective-sphere model.
Then, following Wang \emph{et al.}, we determine the fractal dimension by fitting Eq.~(\ref{eq:main_scaling}) to the orientation-averaged values of $n_{eff}$ and $a_{eff}$ for each aggregate.
For clarity, we henceforth use $D_{f,i}$ to denote input fractal dimensions and $D_{f,h}$ to denote fractal dimensions determined from fitting Eq.~(\ref{eq:main_scaling}).

\begin{figure}[]
\centering
\includegraphics[width=\textwidth]{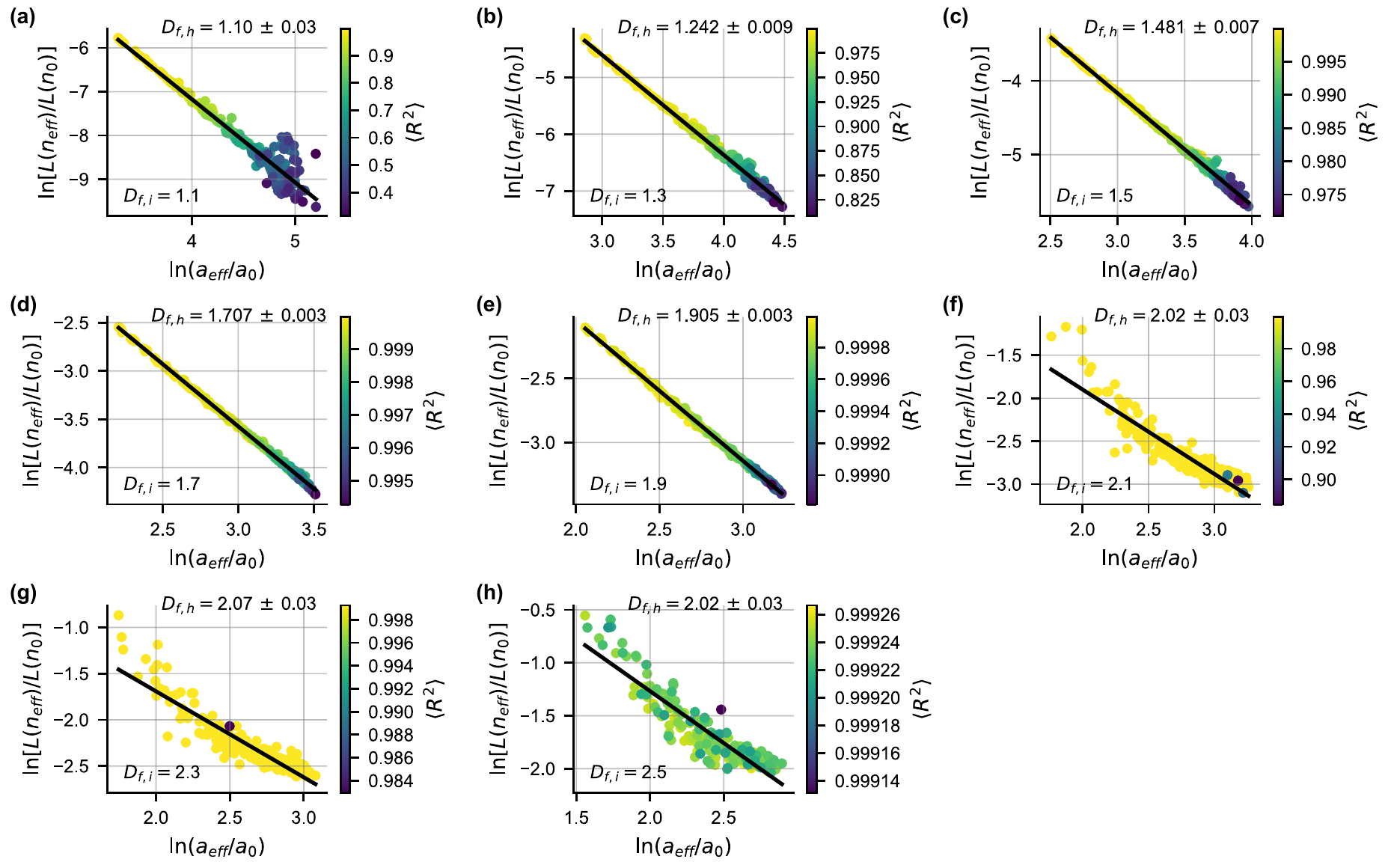}
\caption{\label{fig:distros_r05}
Distributions of orientation-averaged effective refractive indices $n_{eff}$ and effective radii $a_{eff}$ for aggregates composed of spheres with radius $a_0=\SI{5}{\nano\meter}$.
(a) $D_{f,i} = 1.1$; (b) $D_{f,i} = 1.3$; (c) $D_{f,i} = 1.5$; (d) $D_{f,i} = 1.7$; (e) $D_{f,i} = 1.9$; (f) $D_{f,i} = 2.1$; (g) $D_{f,i} = 2.3$; (h) $D_{f,i} = 2.5$.
$L(n)$ denotes the Lorentz-Lorenz factor [Eq.~(\ref{eq:lorentz_lorenz})].
Solid lines are fits of Eq.~(\ref{eq:main_scaling}) and $D_{f,h}$ denotes the values of the fractal dimension determined from those fits.
The color bars show the orientation-averaged values of the coefficient of determination $R^2$ obtained from fitting effective-sphere models to exact holograms.
}
\end{figure}

Figure \ref{fig:distros_r05} shows orientation-averaged values of $n_{eff}$ and $a_{eff}$ for aggregates composed of primary spheres with radius $a_0 = \SI{5}{\nano\meter}$.
The color of each dot indicates $\langle R^2\rangle$, the orientation-averaged value of the $R^2$ coefficient of determination.
The solid lines are best fits of the scaling relation Eq.~(\ref{eq:main_scaling}) to the points, and the fractal dimensions $D_{f,h}$ determined from those fits are shown.
The uncertainties in $D_{f,h}$ are determined from the fits of Eq.~(\ref{eq:main_scaling}) but are underestimates since they neglect the spread in the values of $a_{eff}$. 

For $1.1 \le D_{f,i} \le 1.9$, we find excellent agreement between the data points for each aggregate geometry and Eq.~(\ref{eq:main_scaling}). 
For these values of $D_{f,i}$, the agreement between $D_{f,h}$ and $D_{f,i}$ is 4.5\% or better.
$\langle R^2\rangle$ generally decreases as $a_{eff}$ increases. 
This is particularly noticeable for the relatively linear aggregates with $D_{f,i}=1.1$ in Fig.~\ref{fig:distros_r05}(a). 
For large, nearly-linear aggregates, the agreement between the exact holograms and an effective-sphere model can be poorer when the long axis of the aggregate is nearly perpendicular to the optical axis.

For $2.1\le D_{f,i} \le 2.5$, however, the agreement between the shape of the distribution of points and Eq.~(\ref{eq:main_scaling}) is less good [Fig.~\ref{fig:distros_r05}(f)--(h)]. 
For the smallest clusters (with the lowest $a_{eff}$), the values of $L(n_{eff})/L(n_0)$ lie above the prediction of Eq.~(\ref{eq:main_scaling}).
The points with the lowest $a_{eff}$ values skew the fits of Eq.~(\ref{eq:main_scaling}) and affect the determination of $D_{f,h}$. 
For the aggregates with $D_{f,i}=2.5$, the value of $D_{f,h}$ is nearly 20\% lower than the input value.
We attribute this behavior to multiple scattering, a claim we explore further in Sec.~\ref{subsec:high_df}.

\begin{figure}[]
\centering
\includegraphics[width=\textwidth]{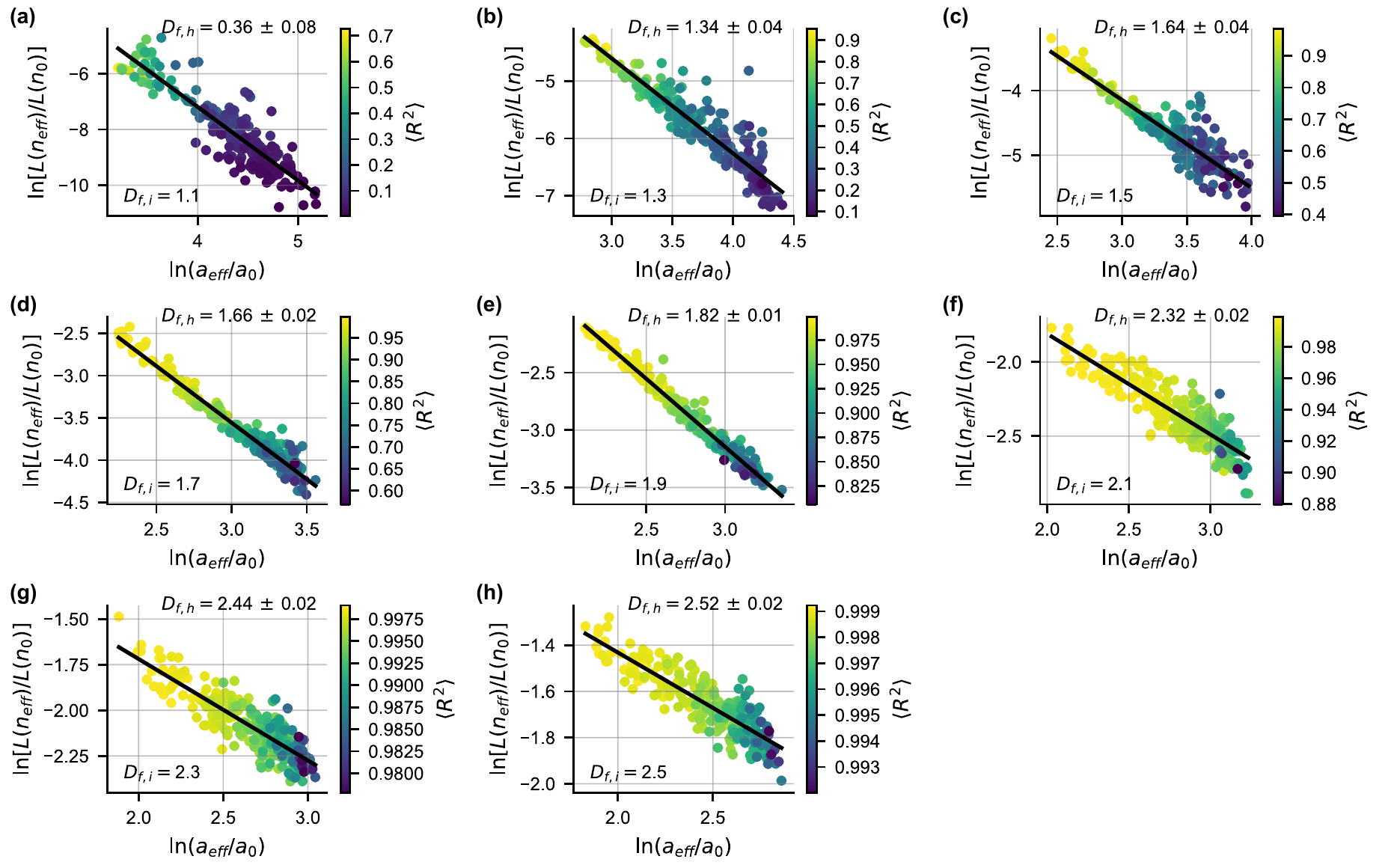}
\caption{\label{fig:distros_r20}
As in Figure \ref{fig:distros_r05}, but for aggregates composed of primary spheres with radius $a_0 = \SI{20}{\nano\meter}$.}
\end{figure}

\begin{figure}[]
\centering
\includegraphics[width=\textwidth]{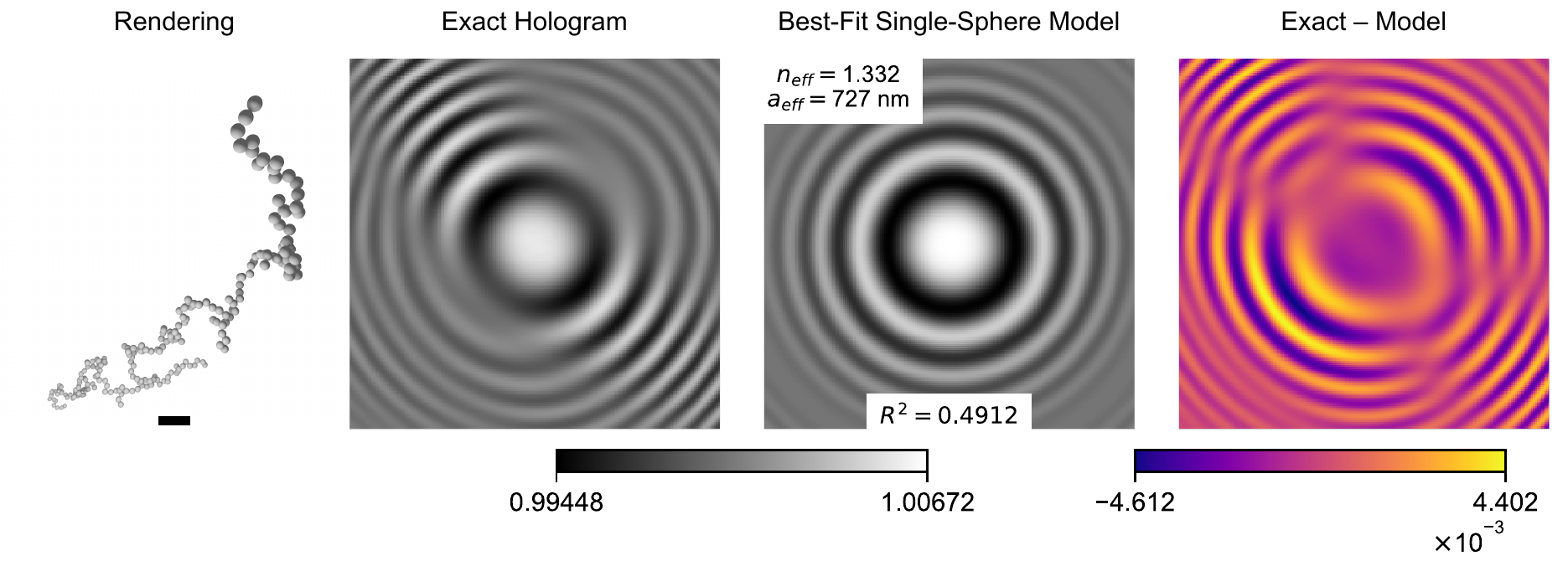}
\caption{\label{fig:bad_fit}
Aggregate structure, exact hologram, best-fit effective-sphere hologram, and residuals for an aggregate with $D_{f,i} = 1.3$ composed of $N=240$ 20-nm-radius spheres with $n_0=1.59$. 
The effective index $n_{eff}$, the effective radius $a_{eff}$ and the $R^2$ coefficient of determination for the fit are also shown. 
Scale bar: 200 nm.
}
\end{figure}

We observe similar results for aggregates composed of primary spheres with radius $a_0=\SI{20}{\nano\meter}$ (Fig.~\ref{fig:distros_r20}).
For aggregates with $D_{f,i}\ge1.3$, $D_{f,h}$ agrees with $D_{f,i}$ to 10.5\% or better.
In addition, we do not observe significant deviations from the scaling predicted by Eq.~(\ref{eq:main_scaling}) for small aggregates with $D_{f,i}\ge 2.1$.
However, at low $D_{f,i}$, the quality of the effective-sphere hologram fits becomes poor, as indicated by the values of $\langle R^2\rangle \ll 1$. 
For $D_{f,i}=1.1$ [Fig.~\ref{fig:distros_r20}(a)], the quality of the fits is so poor that Eq.~(\ref{eq:main_scaling}) does not correctly determine the fractal dimension.
This is because low-$D_f$ (and hence, highly elongated) aggregates composed of 20-nm-radius spheres have a size approaching or exceeding the wavelength of light.
Consequently, the exact holograms of these aggregates exhibit pronounced deviations from axisymmetry. 
Figure \ref{fig:bad_fit} shows holograms of an aggregate with $D_{f,i}=1.3$ consisting of $N=240$ spheres.
While the central bright spot in the exact hologram is well-captured by the effective-sphere fit, the exact hologram exhibits pronounced diagonal bands due to interference between the waves scattered by different spheres within the aggregate \cite{fung_imaging_2012}.

\begin{figure}[]
\centering
\includegraphics[width=\textwidth]{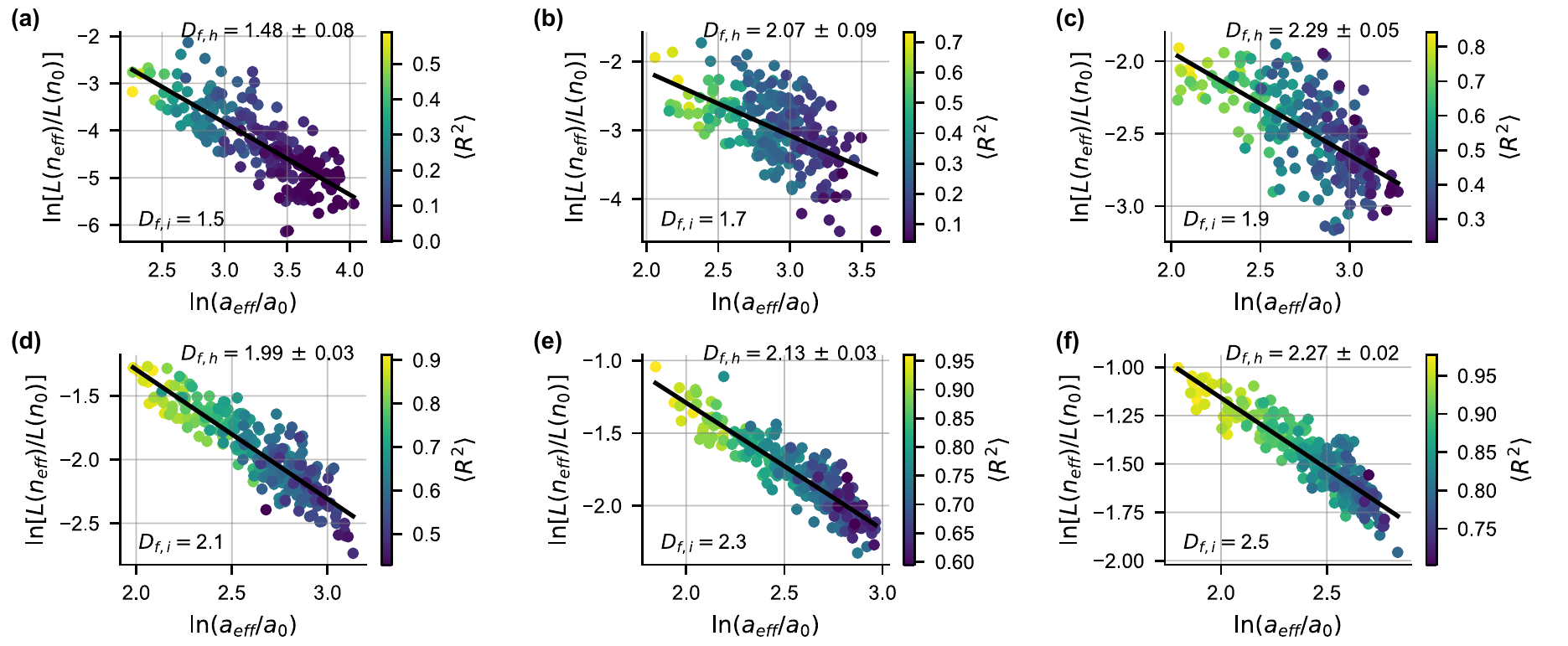}
\caption{\label{fig:distros_r80}
As in Figure \ref{fig:distros_r05}, but for aggregates composed of primary spheres with radius $a_0 = \SI{80}{\nano\meter}$.
The input fractal dimensions $D_{f,i}$ are (a) $D_{f,i} = 1.5$; (b) $D_{f,i} = 1.7$; (c) $D_{f,i} = 1.9$; (d) $D_{f,i} = 2.1$; (e) $D_{f,i} = 2.3$; (f) $D_{f,i} = 2.5$.
}
\end{figure}

For aggregates composed of primary spheres of radius $a_0=\SI{80}{\nano\meter}$, the discrepancies between the exact holograms and the effective-sphere holograms are so large for low-$D_{f,i}$ aggregates that most of the effective-sphere fits do not successfully converge.
For this primary particle size, we therefore only consider aggregates with $1.5 \le D_{f,i} \le 2.5$ (Fig.~\ref{fig:distros_r80}).
Even at $D_{f,i}=1.5$, all the values of $\langle R^2\rangle$ are smaller than 0.6.
Only for $D_{f,i}\ge 2.1$ do the fits of Eq.~(\ref{eq:main_scaling}) determine $D_{f,h}$ to within 10\% of the input values.

\begin{figure}[]
\centering
\includegraphics[]{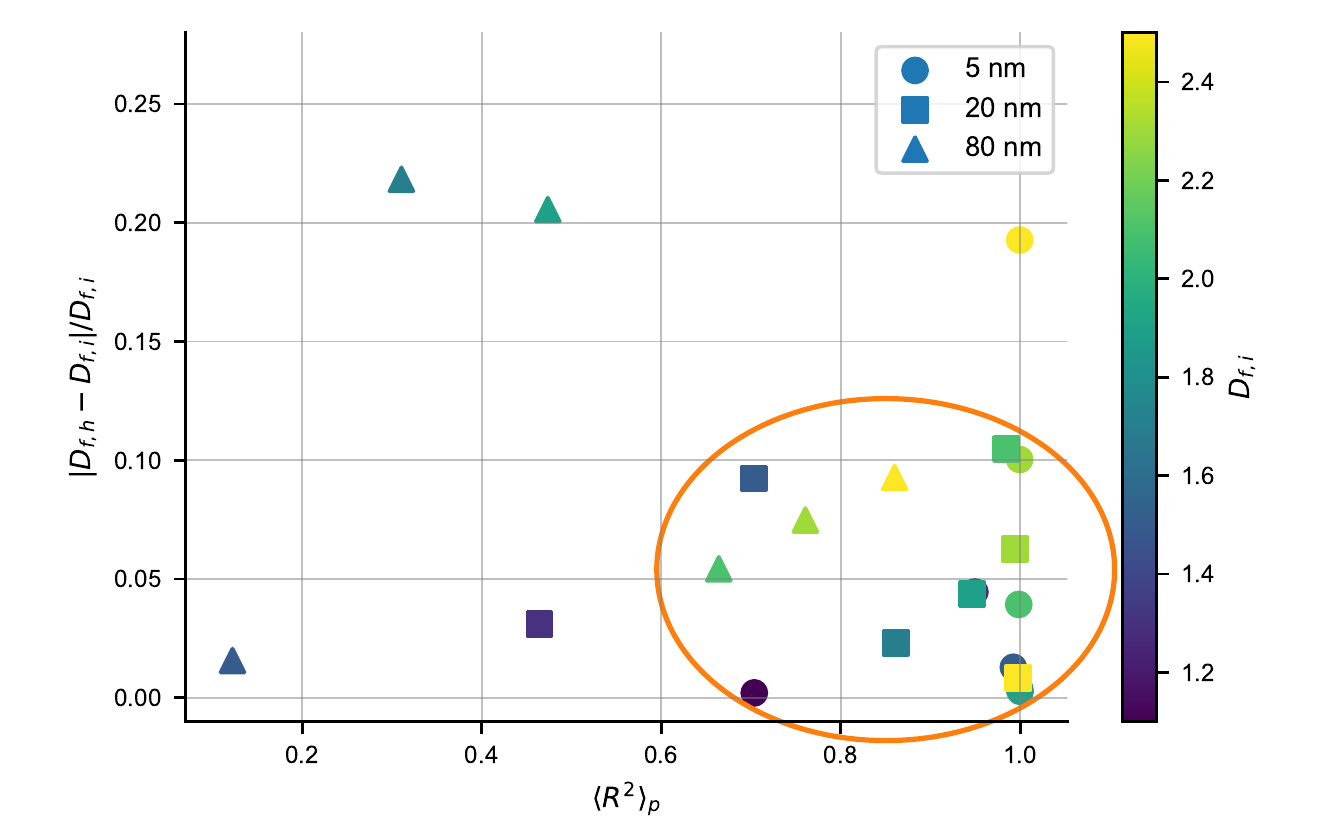}
\caption{\label{fig:r2_summary} Fractional difference between fractal dimensions $D_{f,h}$ determined from Eq.~(\ref{eq:main_scaling}) and input values $D_{f,i}$ as a function of the population-averaged coefficient of determination $\langle R^2\rangle_p$.
Circles: aggregates with $a_0=\SI{5}{\nano\meter}$; squares: aggregates with $a_0=\SI{20}{\nano\meter}$; triangles: aggregates with $a_0=\SI{80}{\nano\meter}$. 
The color bar shows $D_{f,i}$. 
The circled points exhibit both small ($\lessapprox 10\%$) fractional differences between $D_{f,h}$ and $D_{f,i}$ and large ($>0.6$) values of $\langle R^2\rangle_p$.
}
\end{figure}

For each of the primary particle radii we have considered, the scaling relationship in Eq.~(\ref{eq:main_scaling}) accurately determines the fractal dimension for some but not all of the values of $D_{f,i}$. 
In order to draw conclusions about the conditions under which Eq.~(\ref{eq:main_scaling}) is useful, we summarize our results on $D_{f,h}$ in Fig.~\ref{fig:r2_summary}.
Here we show the fractional difference between $D_{f,h}$ and $D_{f,i}$ as a function of $\langle R^2\rangle_p$, the population-averaged value of $R^2$ for all hologram fits at a given $D_{f,i}$ (e.g., all of the points within one of the panels of Fig.~\ref{fig:distros_r05}, \ref{fig:distros_r20}, and \ref{fig:distros_r80}).
Several features are apparent in Fig.~\ref{fig:r2_summary}.
The points inside the orange oval have both a small fractional difference ($\lessapprox 10\%$) between $D_{f,h}$ and $D_{f,i}$ as well as high values ($>0.6$) of the population-averaged coefficient of determination $\langle R^2\rangle_p$.
There are also four points (corresponding to low-$D_{f,i}$ aggregates composed of 20- or 80-nm-radius primary particles) for which the quality of the fits is poor: $\langle R^2\rangle_p< 0.5$.
For two of these points, there is a large ($>20\%$) discrepancy between $D_{f,h}$ and $D_{f,i}$.
While for the other two points, the discrepancy between $D_{f,h}$ and $D_{f,i}$ is less than 4\%, these results suggest that Eq.~(\ref{eq:main_scaling}) is unreliable when $\langle R^2\rangle_p \lesssim 0.6$.

There is one other significant feature in Fig.~\ref{fig:r2_summary}. For aggregates with $D_{f,i}=2.5$ composed of 5-nm-radius spheres, even though $\langle R^2\rangle_p = 0.9992$, the fractional discrepancy between $D_{f,i}$ and $D_{f,h}$ is 19.3\%.
This case is shown in Fig.~\ref{fig:distros_r05}(h).
We next consider the origin of this discrepancy, which we attribute to the breakdown of the Maxwell Garnett effective-medium theory for high-$D_f$ aggregates composed of small primary spheres due to multiple scattering.

\subsection{Breakdown of effective-medium theory at high $D_f$} \label{subsec:high_df}

The results for aggregates with $D_{f,i}\ge2.1$ composed of 5-nm-radius primary particles in Fig.~\ref{fig:distros_r05}
deviate significantly from the scaling predictions of Eq.~(\ref{eq:main_scaling}) for the smallest values of $a_{eff}$.
Theoretical considerations along with additional computations suggest that these deviations are due to the breakdown of the Maxwell Garnett effective-medium theory used to derive Eq.~(\ref{eq:main_scaling}) as a result of multiple scattering.

The Maxwell Garnett theory assumes that an optical medium (here, the combination of a fractal aggregate and the medium surrounding it) can be modeled as consisting of dipolar inclusions within a homogeneous background.
The volume fraction $\phi$ of the inclusions is assumed to be low, and the density of inclusions is assumed to be constant.
As Wang \emph{et al.} point out, one problem with applying the Maxwell Garnett theory to fractal aggregates is that the particles are not dispersed uniformly within an aggregate: there are more particles near the center of the aggregate than towards the edges \cite{wang_holographic_2016}.
(We note, as do Wang \emph{et al.}, that a more realistic effective-sphere model for a fractal aggregate could incorporate a radially-varying refractive index, as in the work of Khlebtsov \cite{khlebtsov_optics_1993}.)
More importantly, all the inclusions are assumed to experience the same electric field, which requires that multiple scattering be negligible.

A scaling argument due to Lu and Sorensen \cite{lu_depolarized_1994, sorensen_light_2001} suggests that multiple scattering may be non-negligible for low-$N$ aggregates composed of small primary spheres in the Rayleigh limit, for which $ka_0\ll1$.
In this limit, each sphere scatters like a dipole.
At distances $r$ from a dipole such that $kr<1$, the near-field term in the scattered electric field (which is proportional to $(kr)^{-3}$) dominates over the usual far-field term (which is proportional to $(kr)^{-1}$).
For aggregates small enough that $kr_{ij}<1$ for every interparticle distance $r_{ij}$ within the aggregate, every primary particle is within
the near field of every other particle.
Lu and Sorensen consider the amplitude of the double-scattered electric field from an aggregate under this assumption and show that the ratio of the double-scattered to single-scattered fields is proportional to $N^{1-3/D}$. 
Crucially, this ratio increases as $N$ decreases. 
This explains why we observe deviations from Eq.~(\ref{eq:main_scaling}) for low-$N$ aggregates with $D_{f,i}\ge 2.1$ when $a_0 = \SI{5}{\nano\meter}$.
(When $a_0=\SI{5}{\nano\meter}$, $ka_0 = 0.093$, and for an aggregate with $D_{f,i}=2.5$ composed of $N=100$ spheres, $kR_g = 0.587$.)
We do not observe similar deviations from Eq.~(\ref{eq:main_scaling}) for aggregates with $a_0=\SI{5}{\nano\meter}$ and $D_{f,i}\le1.9$ or aggregates of any $D_{f,i}$ composed of larger primary particles since it is not the case that every primary particle in such aggregates is in the near field of every other particle.

\begin{figure}[]
\centering
\includegraphics[]{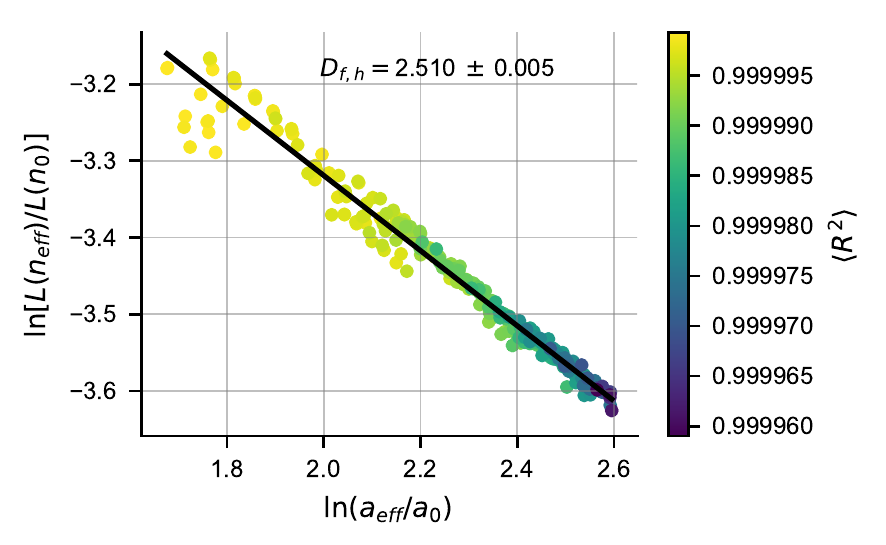}
\caption{\label{fig:low_n} Orientation-averaged $n_{eff}$ and $a_{eff}$ for aggregates with input fractal dimension $D_{f,i}=2.5$ composed of 5-nm-radius spheres with refractive index $n_0 = 1.356$ in water ($n_{med}=1.33$). 
Aggregate geometries are identical to those in Fig.~\ref{fig:distros_r05}(h). 
The color bar shows the orientation-averaged value of the $R^2$ coefficient of determination for each aggregate, and the solid line is a best fit of Eq.~(\ref{eq:main_scaling}). }
\end{figure}

Further evidence suggesting that multiple scattering is significant for high-$D_{f,i}$ aggregates of 5-nm-radius particles comes from simulations in which we decrease the effects of multiple scattering by decreasing the refractive index of the primary particles.
Figure \ref{fig:low_n} shows $n_{eff}$ and $a_{eff}$ for aggregates with $D_{f,i}=2.5$ and $a_0=\SI{5}{\nano\meter}$, but with $n_0 = 1.356$. 
Here, the refractive index difference $n_0 - n_{med}$ is an order of magnitude smaller than for the polystyrene spheres ($n_0=1.59$) previously considered. 
While the aggregate geometries and orientations in Fig.~\ref{fig:low_n} are \emph{identical} to those in Fig.~\ref{fig:distros_r05}(h),
the agreement between the data and Eq.~(\ref{eq:main_scaling}) is much better, and $D_{f,h}$ is determined with an accuracy of 0.4\%.

\subsection{Effective-sphere approach for digital holography of fractal aggregates}
Why does the effective-sphere approach for analyzing digital holograms of fractal aggregates succeed when it does?
The scaling approach of Sorensen \cite{sorensen_light_2001} is a useful way to consider this question.
Sorensen argues that there are two important regimes in the functional behavior of the structure factor $S(q)$, which is proportional to the scattered intensity, for scattering by a fractal aggregate of fractal dimension $D_f$ and radius of gyration $R_g$:
\be \label{eq:sorensen_scaling}
S(q) = \begin{cases} 
1 & qR_g < 1 \\
C(qR_g)^{-D_f} & qR_g > 1.
\end{cases}
\ee
Here $C$ is a proportionality constant and $q$ is the scattering wavenumber, defined by $q \equiv (4 \pi n_{med}/\lambda) \sin(\theta/2)$, where $\theta$ is the scattering angle.
The key idea is that near the forward direction (small $q$), the scattered intensity is insensitive to $D_f$, while at larger angles the angular dependence encodes $D_f$.

Our results suggest that the effective-sphere approach succeeds only when $qR_g \lessapprox 1$.
In our simulations, the maximum value of $q$ (at the corners of the detector) is given by $q_{max} = \SI{7.45}{\micro\meter^{-1}}$.
For given values of $a_0$, $D_f$, $k_f$, and $N$, we can determine $R_g$ using Eq.~(\ref{eq:df_defn}).
For aggregates with $a_0=\SI{5}{\nano\meter}$, $q_{max}R_g$ is generally of order 1 or smaller. 
For instance, for an aggregate of 400 spheres with $a_0 = \SI{5}{\nano\meter}$, $D_f = 1.5$, and $k_f=1.4$, $q_{max}R_g = 1.62$.
But when $a_0 = \SI{20}{\nano\meter}$, $q_{max}R_g = 6.46$ for the same aggregate, and when $a_0=\SI{80}{\nano\meter}$, $q_{max}R_g = 25.9$.
As we have seen, for aggregates composed of larger primary particles where $q_{max}R_g \gg 1$, the quality of the effective-sphere fits tends to be poor.
The poor quality of the fits primarily arises from the lack of axial symmetry in the holograms, but Eq.~(\ref{eq:sorensen_scaling}) suggests that the angular dependence of the scattered intensity (which affects the envelope of the hologram fringe pattern) is also inconsistent with scattering by a uniform sphere.
In the Rayleigh-Gans approximation for scattering by a sphere, the envelope of the scattered intensity scales as $q^{-4}$, in contrast to the $q^{-D_f}$ scaling for a fractal aggregate \cite{sorensen_light_2001}.

Because the scattered intensity does not depend on $D_f$ when $qR_g<1$, in the approach analyzed here, $D_f$ cannot be determined from a single hologram of an aggregate.
But since holograms capture the scattered field at small scattering angles (including the forward direction), it follows from the optical theorem that holograms probe the extinction cross section of scatterers \cite{berg_using_2014}.
For the non-absorbing scatterers considered here, the extinction cross section is equal to the scattering cross section.
The data in Figs.~\ref{fig:distros_r05}, \ref{fig:distros_r20}, and \ref{fig:distros_r80} thus encode how the scattering cross section scales with the size of an aggregate.
Within the Rayleigh-Gans approximation, the scattering cross section for fractal aggregates scales with $D_f$ and $R_g$ \cite{dobbins_absorption_1991, sorensen_light_2001}.
This dependence enables us to retrieve $D_f$.

\section{Conclusions}
It is relatively straightforward to incorporate holographic microscopy into an existing optical setup, and the effective-sphere approach for characterizing fractal aggregates introduced by Wang \emph{et al.} may be of interest to other experimenters.
We have evaluated this approach by computing exact holograms of fractal aggregates using a multisphere superposition code.
Our simulations suggest criteria that may help experimenters assess whether or not the fractal dimensions they determine using this approach are reliable to within $\sim10$\% or better.
First, the effects of multiple scattering should be negligible.
We have shown how the effects of multiple scattering may be observed in deviations from the Maxwell Garnett scaling relationship [Eq.~(\ref{eq:main_scaling})] for aggregates for which $kR_g\lessapprox 1$.
Second, as shown in Fig.~\ref{fig:r2_summary}, the population-averaged values of $R^2$ should be sufficiently large ($\langle R^2\rangle_p \gtrapprox 0.6$) to indicate that the effective-sphere models reliably describe the holograms.
These criteria are useful because estimates of $kR_g$ and $\langle R^2\rangle_p$ can be determined from the experimental holograms alone.
No \emph{a priori} knowledge of the radius and refractive index of the primary particles is necessary; in many experimental applications (such as the protein aggregates considered by Wang \emph{et al.}), the optical properties of the primary particles are unknown.

Several further extensions of this work are possible. First, in this work we have only carefully analyzed the determination of the population-averaged values of $D_f$. While Wang \emph{et al.} assumed that the fractal prefactor $k_f=1$, it should in principle be possible to determine the fractal prefactor $k_f$ using Eq.~(\ref{eq:main_scaling}) as well.
Second, our simulations assume that \emph{every} aggregate in a population has the same $D_f$ and $k_f$, and that the primary particles are monodisperse.
These assumptions are unlikely to be satisfied for naturally-occurring aggregates.
Future work might thus explore the effects of particle polydispersity and variations in $D_f$.
Finally, this work has only considered aggregates of optically non-absorbing particles. 
We intend to extend these results to absorbing particles in the future, as well as to non-absorbing primary particles with other refractive indices.
Nonetheless, our work confirms the findings of Wang and co-workers that holographic microscopy is an easy-to-use and effective tool for characterizing fractal aggregates.

\section*{Acknowledgements}
S.~H.~was supported by the Wellesley College IBM Faculty Research Fund for Science and Mathematics.


\bibliography{ms}

\end{document}